\documentclass[%
 reprint,
nofootinbib,
 amsmath,amssymb,
 aip,
floatfix,
longbibliography,
]{revtex4-2}

\usepackage{graphicx}% Include figure files
\usepackage{subcaption}
\usepackage{dcolumn}% Align table columns on decimal point
\usepackage{bm}% bold math
\usepackage[english]{babel}
\usepackage{upgreek}
\usepackage{ifthen}
\usepackage{subcaption}
\usepackage{comment}
\usepackage{tikz}
\usepackage{pgfplots}
\pgfplotsset{compat=1.17}

\usepackage{diagbox}

\newcommand{\bs}[1]{\boldsymbol{#1}}

\newcommand{\paren}[1]{\left(#1\right)}

\newcommand{\wt}{\widetilde}

\newcommand{\avg}[1]{\langle #1 \rangle}

\newcommand{\mr}[1]{\mathrm{#1}}

\newcommand{\vth}{v_\mr{th}}

\newcommand{\sfe}{SFRE}

\newcommand{\Tign}{T_\mr{ign}}
\newcommand{\chiTKE}{\chi_\mr{turb}}
\newcommand{\tauE}{\tau}

\usepackage{mathptmx}
\usepackage{etoolbox}
\usepackage{ragged2e}

\usepackage[font=small]{caption}

\makeatletter
\def\@email#1#2{%
 \endgroup
 \patchcmd{\titleblock@produce}
  {\frontmatter@RRAPformat}
  {\frontmatter@RRAPformat{\produce@RRAP{*#1\href{mailto:#2}{#2}}}\frontmatter@RRAPformat}
  {}{}
}%
\makeatother
\begin{document}

%\preprint{APS/123-QED}
\preprint{AIP/123-QED}

\title{An ignition criterion for inertial fusion boosted by microturbulence}

\author{Henry Fetsch}
 \email{hfetsch@princeton.edu}
\author{Nathaniel J. Fisch}%
\affiliation{Department of Astrophysical Sciences, Princeton University, Princeton, NJ}

\date{\today}

%COAUTHOR QUESTIONS
% Title...
% Abbreviation (SFRE etc...)

\begin{abstract}
Turbulence enhances fusion reactivity, enabling ignition at lower temperature. A modified Lawson-like ignition criterion is derived for inertially confined plasmas harboring turbulent kinetic energy. Remarkably, if small-scale turbulence is driven in the hot spot while avoiding mixing at the boundary, less energy is required to ignite a target. 
The optimal length scale for hot-spot turbulence is quantified, typically lying in the micron range.
%Turbulent eddy sizes, generally on the micron scale, are identified where driving of such flows is most advantageous.
\end{abstract}

\maketitle

Thermonuclear ignition requires that fuel be driven to high temperature and density and be sufficiently well confined that heating by fusion products exceeds energy losses. 
The classic Lawson criterion \cite{Lawson_1957} quantifies this requirement. 
Inertial confinement fusion (ICF) involves the complication that the hot fuel is typically disassembling even while igniting, thus losing energy to mechanical work in addition to the radiation and transport present in magnetic confinement devices. 
The ICF community has therefore developed a variety of generalized Lawson criteria (GLC), some based on power balance \cite{Lindl_et_2018,Zhou_Betti_2008,Betti_et_2010,Christopherson_et_2018,Zylstra_et_2022}, on other thermodynamic properties of the hot spot \cite{Hurricane_et_2021,Hurricane_et_2024,Christopherson_et_2020}, or on related criteria such as the positivity of the second time-derivative of the hot-spot temperature %$(d^2T/dt^2 > 0)$ 
\cite{Lindl_et_2018}. 

%Generally, ignition criteria can be formulated either in terms of internal hot-spot parameters or in terms of direct experimental observables.
Alternatively, ignition may simply be defined as fusion gain exceeding unity \cite{ICF_collab_2024}.  
In ICF, maximizing gain requires that compression expend minimal energy per unit of fusion yield. 
Other than the necessary heating of fuel ions, any process that increases energy losses, reduces confinement time, or consumes compression energy degrades the gain of ICF experiments. 
Residual kinetic energy (RKE), the portion of the implosion energy that remains in flows rather than thermalizing, is detrimental in all of these respects. 
Success in reducing RKE has contributed to improvements in ICF performance and the achievement of ignition\cite{Kritcher_et_2024,Kritcher_et_2024,Hurricane_et_2024,Rosen_2024,ICF_collab_2024}.

However, this is not the full story. 
As a fast ion travels through a plasma containing small-scale turbulent fluctuations, its peculiar velocity varies with respect to the flow velocities of local fluid elements. A particle near the thermal bulk in one region can end up further out on the tail if it travels to another region with different flow. The result is a broadening of the tail of the ion distribution function and, therefore, an enhancement to fusion reactivity \cite{Fetsch_Fisch_2024}. This recently introduced shear flow reactivity enhancement (\sfe)  can multiply reactivity severalfold under ICF-relevant conditions \cite{Fetsch_Fisch_2024,Fetsch_Fisch_formula}. 
Because fusion reactivity is determined by fast ions, whose mean free paths are much longer than those of thermal particles, a large reactivity enhancement is possible even when flow gradients are weak and so viscous dissipation is relatively slow. 
It follows that some turbulent kinetic energy (TKE) might, in fact, be advantageous in achieving ignition. 
%The question of optimal partition between thermal and turbulent energy was posed in Ref.\citenum{Fetsch_Fisch_2024} but has so far not been answered.

Such an advantage is not captured in the standard Lawson criterion $p \tau > F(T)$ (for pressure $p$, confinement time $\tau$, and a function $F$ of the temperature $T$) \cite{Lawson_1957,Betti_et_2010}, which is agnostic of TKE. 
Based on the ignition parameter $\chi$ of Betti \textit{et al.} \cite{Betti_et_2010}, defined such that ignition means ${\chi > 1}$, this work introduces a new ignition criterion for turbulent, inertially confined plasmas. 
Our modified ignition parameter $\chiTKE$ accounts for the \sfe~and is a functional of the turbulent energy spectrum in the hot spot. 
Remarkably, TKE on short length scales %(``microturbulence'') 
increases $\chiTKE$ more efficiently than thermal energy; TKE on longer scales is detrimental, in keeping with the conventional picture. 
In some cases, an otherwise non-igniting target will ignite if a portion of the implosion energy is redirected into TKE. This class of implosions, characterized by ${\chiTKE > 1 > \chi}$, represents a new ICF regime enabled by turbulence. 

\begin{figure}
        \centering
        \begin{tikzpicture}[scale=0.42, every node/.style={scale=0.5}, line cap=round, line join=round]
            % Define colors
            \colorlet{hotfluid}{red!90!black}
            \colorlet{coolfluid}{red!80!blue!70!black!80}
            \colorlet{coldshell}{black!90!blue!90!red!90}

            \colorlet{radiation}{blue!90!red!90!white!90}
            
            %surrounding cold shell
            \fill[coldshell] (0,0) circle (4.25cm);
            % First circle - hot turbulent fluid with swirls
            \fill[hotfluid] (0,0) circle (3.5cm);

            % Add outward-pointing arrows around the inner shell to show expansion
            \foreach \angle in {0, 30, 60, 90, 120, 150, 180, 210, 240, 270, 300, 330} {
              \draw[->, thick, white] ({3.5*cos(\angle)},{3.5*sin(\angle)}) -- ({4.1*cos(\angle)},{4.1*sin(\angle)});
            }

            % Add wavy lines leaving the inner circle to represent radiation losses
            \foreach \angle in {15, 75, 135, 195, 255, 315} {
              \draw[radiation, line width=0.06cm, decorate, decoration={snake, amplitude=0.1cm, segment length=0.6cm}, ->]
                      ({2.5*cos(\angle)},{2.5*sin(\angle)}) -- ({4.8*cos(\angle)},{4.8*sin(\angle)});
            }

            % Add swirling eddies to the first circle - with better spacing and consistent orientation
            \foreach \x/\y in {
                -2.0/-1.0, 
                -0.5/2.0, 
                %1.7/1.2, 
                1.6/0.3, 
                %1.8/-1.3, 
                -0.1/-2.0
                %-2.0/1.2, 
                %1.0/0.0,
                %0.0/0.0
            } {
                \draw[white, line width=0.05cm] 
                    (\x,\y) arc[start angle=145, end angle=-135, radius=0.5cm];
                    \draw[white, line width=0.05cm, ->] 
                    (\x, \y) --
                    ({\x - 0.1*sin(145)}, {\y + 0.1*cos(145)});
                    %({\x + 0.5*cos(-135) - 0.5*cos(145)},{\y + 0.5*sin(-135) - 0.5*sin(145)}) -- ({\x + 0.5*cos(-135) + 0.5*sqrt(2)},{\y + 0.5*sin(-135) + 0.5*sqrt(2)});
            }
            
            \fill[coldshell] (10,0) circle (4.25cm);
            % Second circle - cooler fluid - with fewer swirls
            \fill[coolfluid] (10,0) circle (3.5cm);

            % Add outward-pointing arrows around the inner shell to show expansion
            \foreach \angle in {0, 30, 60, 90, 120, 150, 180, 210, 240, 270, 300, 330} {
              \draw[->, thick, white] ({10+3.5*cos(\angle)},{3.5*sin(\angle)}) -- ({10+4.1*cos(\angle)},{4.1*sin(\angle)});
            }

            % Add wavy lines leaving the inner circle to represent radiation losses
            \foreach \angle in {15, 135, 255} {
              \draw[radiation, line width=0.06cm, decorate, decoration={snake, amplitude=0.1cm, segment length=0.6cm}, ->]
                      ({10+2.5*cos(\angle)},{2.5*sin(\angle)}) -- ({10+4.8*cos(\angle)},{4.8*sin(\angle)});
            }   
            
            % Add medium-sized eddies - fewer, with better spacing and consistent orientation
            \foreach \x/\y in {
                8.2/1.5,
                8.9/2.4,
                9.5/1.0,
                10.2/2.0,
                10.8/1.3,
                11.4/2.0,
                %11.8/1.0,
                7.2/0.4,
                8.2/-1.5,
                %8.8/-2.2,
                9.5/-1.0,
                10.2/-2.0,
                10.8/-1.3,
                %11.4/-2.0,
                11.8/-1.0,
                7.5/-0.5,
                8.7/0.0,
                10.0/0.0,
                11.3/0.0
            } {
                \draw[white, line width=0.04cm] 
                    (\x,\y) arc[start angle=145, end angle=-135, radius=0.4cm];
                    \draw[white, line width=0.04cm, ->] 
                    (\x, \y) --
                    ({\x - 0.08*sin(145)}, {\y + 0.08*cos(145)});
            }
            
            % Set manual text size for the figureExample: sets font size for all nodes below
            \node[above, font=\fontsize{20}{14}\selectfont] at (0,4.75) {High Temp.};
            \node[above, font=\fontsize{20}{14}\selectfont] at (10,4.55) {High TKE};

            %\node[below, font=\fontsize{18}{14}\selectfont] at (0,-3.8) {\textit{High reactivity}};
            %\node[below, font=\fontsize{18}{14}\selectfont] at (10,-3.8) {\textit{Conventionally: low reactivity}};
            
            % Updated energy transfer arrow position
            %\draw[->, very thick] (2.5,-2) -- (7.5,-2) node[midway, above] {Energy Transfer};
        \end{tikzpicture}

        \caption{\justifying Replacement of thermal energy with TKE. The left panel shows a conventional ICF hot spot (red) with little internal turbulence; on the right, some hot-spot thermal energy is replaced by TKE. The effective pressure is constant, so the expansion rate into the cold shell (grey) is unchanged. However, the turbulent hot spot emits less radiation (blue arrows). The thermal reactivity of the turbulent hot spot is lower, but the turbulent enhancement can be large.}
       
        \label{fig_TKE_sketch}
  \end{figure}
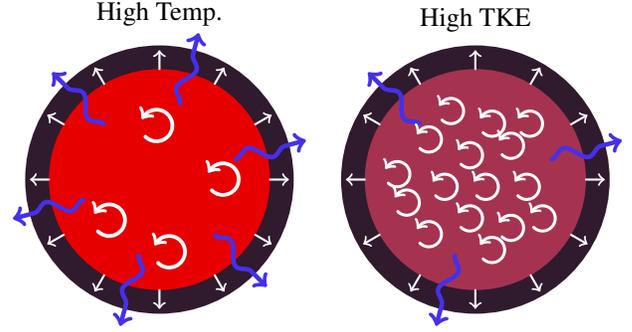

We first consider ignition of a typical isobaric hot spot containing no turbulence, which is assumed for simplicity to be spherically symmetric. After stagnation, the hot spot expands against a surrounding shell of cold, dense fuel (Fig.~\ref{fig_TKE_sketch}, left side). 
Some alpha particles, radiation, and conducted heat escape the hot spot, but a portion of escaping energy is recycled back into the hot spot through ablation of cold fuel \cite{Betti_et_2010,Christopherson_et_2018}. Therefore, as an approximation, we neglect heat conduction. In DT plasma, the self-heating condition is
\begin{equation}
  \label{eq_self_heat_no_turb}
  Q_\alpha > Q_b + Q_w .
\end{equation}
where ${Q_\alpha = f_\alpha p^2 \epsilon_\alpha S(T)}$ is the power due to alpha heating, ${Q_b = f_b p^2 B(T)}$ is the radiated power, and ${Q_w = p/\tauE}$ is the rate of mechanical work. Here, $f_\alpha$ is the fraction of fusion power deposited in the hot spot, $f_b$ is the fraction of radiated power escaping the hot spot, $\epsilon_\alpha$ is the birth energy of alpha particles, and $\tau$ is the confinement time. In terms of the hot-spot central temperature $T$, the function $S(T)$ is defined by
\begin{equation}
  \label{eq_S_def}
  S(T) = p^2 \frac{1}{V}\int_V d^3r \frac{\avg{\sigma v}}{\wt T^2} ,
\end{equation}
where $\avg{\sigma v}$ is the fusion reactivity, $\wt T(\bs r)$ is the spatially varying temperature, and $V$ is the hot-spot volume. 
The pressure, which is taken to be uniform throughout the hot spot, is ${p = 2nT}$, where $n$ is the ion number density. 
We follow previous authors \cite{Betti_et_2010,Zhou_Betti_2008,Christopherson_et_2018} in approximating $S$ and $B$ by power laws ${S(T) \approx S_0 T^{\omega-2}}$ and ${B(T) \approx B_0 T^{\beta-2}}$ for $T$ in keV and with constants ${S_0 \approx 2.78 \times 10^{-21}}$, ${B_0 \approx 8.34\times 10^{-16}}$, ${\omega \approx 3.26}$, and ${\beta \approx 1/2}$. The fit for $S$ is obtained in the region ${4~\mr{keV} < T < 12~\mr{keV}}$ using the temperature profile\cite{Betti_et_2010} ${\wt T(r) \approx (1-r^2/R^2)^{2/5}/(1-0.15r^2/R^2)}$ where $R$ is the hot-spot radius; the model for $B$ assumes that the net rate of radiative losses is roughly uniform throughout the hot spot\cite{Christopherson_et_2018}. Then \eqref{eq_self_heat_no_turb} leads to the ignition condition ${\chi > 1}$, where
\begin{equation}
  \label{eq_chi_def}
  \chi = \paren{f_\alpha \epsilon_\alpha S_0 T^{\omega-2} - f_b B_0 T^{\beta-2}} \tauE p .
\end{equation}

%IMPORTANT NOTE: \cite{Christopherson_et_2018} Appendix A suggests that net radiation losses are approximately constant throughout the hotspot radius. While n^2T^1/2 increases sharply near the boundary, emitted photons have much lower energy and are much more readily absorbed

%The form of $\chi$ in \eqref{eq_chi_def} is not unique -- for example, the radiation losses could be included in the denominator -- but it is physically transparent for the purposes of this work. 
When $\chi > 0$, alpha heating exceeds radiative losses; such a plasma would ignite if it were not expanding. The minimum temperature $T_\mr{ign}$ for static ignition is
\begin{equation}
  \label{eq_T_ign_def}
  \Tign = \paren{\frac{f_b B_0}{f_\alpha \epsilon_\alpha S_0}}^{\frac{1}{\omega - \beta}} 
\end{equation}
(note that $f_\alpha$ and $f_b$ can depend on temperature, so \eqref{eq_T_ign_def} is not an explicit formula for $\Tign$ unless the temperature dependence is negligible). The confinement time can be estimated\cite{Betti_et_2010} as ${\tauE \approx C_E R / v_\mr{imp}}$, where $v_\mr{imp}$ is the implosion velocity and $C_E$ is a constant. 
When $\chi > 1$, alpha heating is strong enough to overcome both radiative losses and expansion, so the plasma ignites. 
This work does not consider subsequent burn propagation, which occurs when $\chi$ is large\cite{Christopherson_et_2020}.

Missing from all ignition criteria to date is an accounting for the direct effect of turbulence on fusion reactivity, the \sfe~ introduced in Ref.~\citenum{Fetsch_Fisch_2024}. Crucially, ions near the Gamow peak, where fusion reactions are most likely to occur, have velocities much greater than the thermal velocity $\vth$. The mean free path of these fast ions is correspondingly much longer than the thermal mean free path $\lambda_\mr{th}$. 
In turbulent plasma, where gradients may be weak on thermal-particle scales but strong on the scales of ions near the Gamow peak, nonlocal kinetic modifications to the ion distribution function can significantly affect fusion reactivity. 
An approximate formula for the reactivity enhancement is derived in Ref.~\citenum{Fetsch_Fisch_formula}. While this formula involves several approximations, and in fact tends to underestimate the reactivity enhancement in high Mach-number flows, it offers a useful quantitative model for the purpose of this work. 
Such a formula is necessary given the enormous computational expense of simulating the \sfe~\textit{ab initio}, requiring both high-resolution viscous fluid simulations and kinetic simulations that accurately model fast-ion dynamics.

\begin{figure} 
  \centering
  \begin{subfigure}[t]{\columnwidth}
    \centering
    \includegraphics[width=\columnwidth]{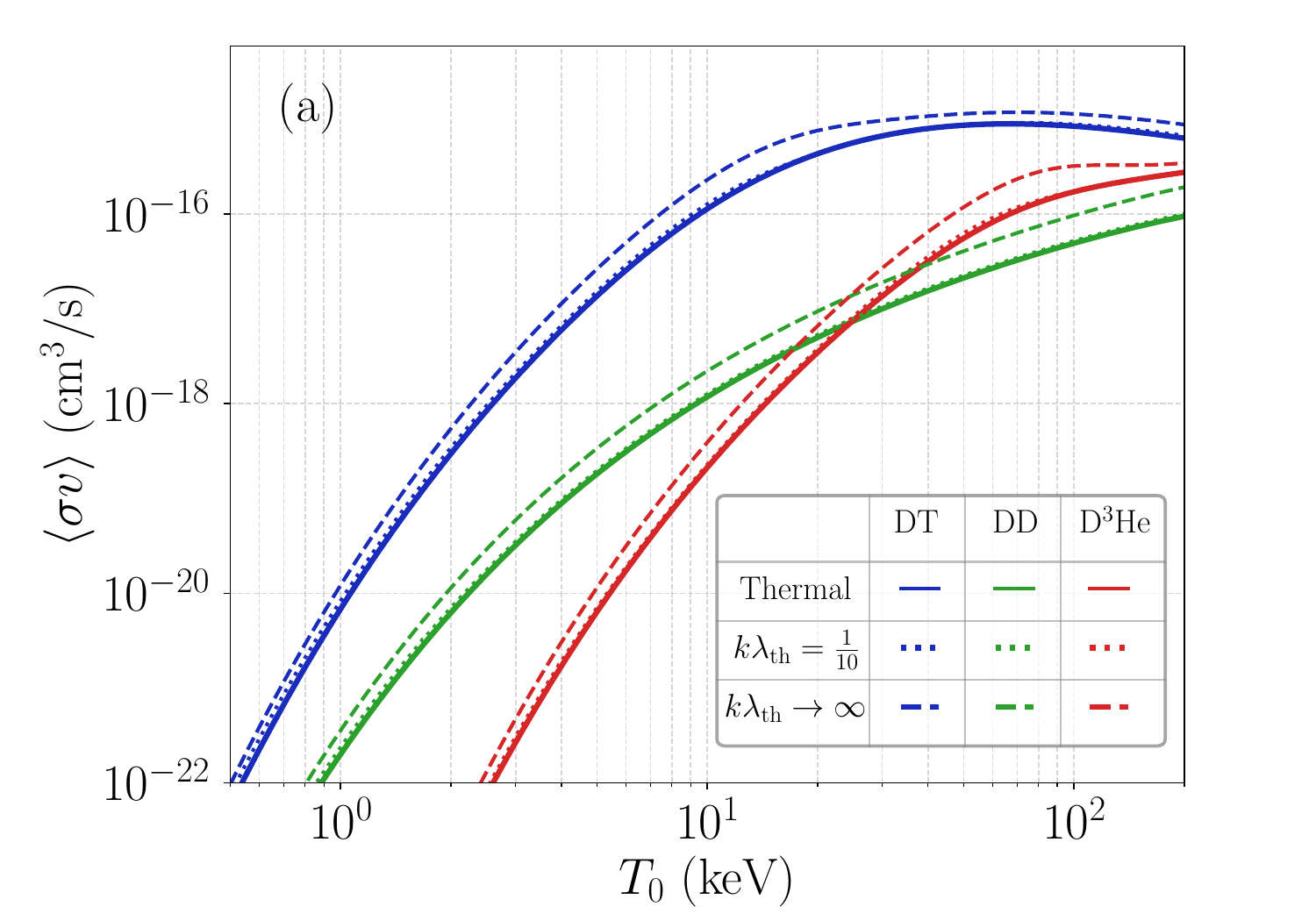}
    \label{fig_reactivity_T0_a}
  \end{subfigure}
  \begin{subfigure}[t]{\columnwidth}
    \centering
    \includegraphics[width=\columnwidth]{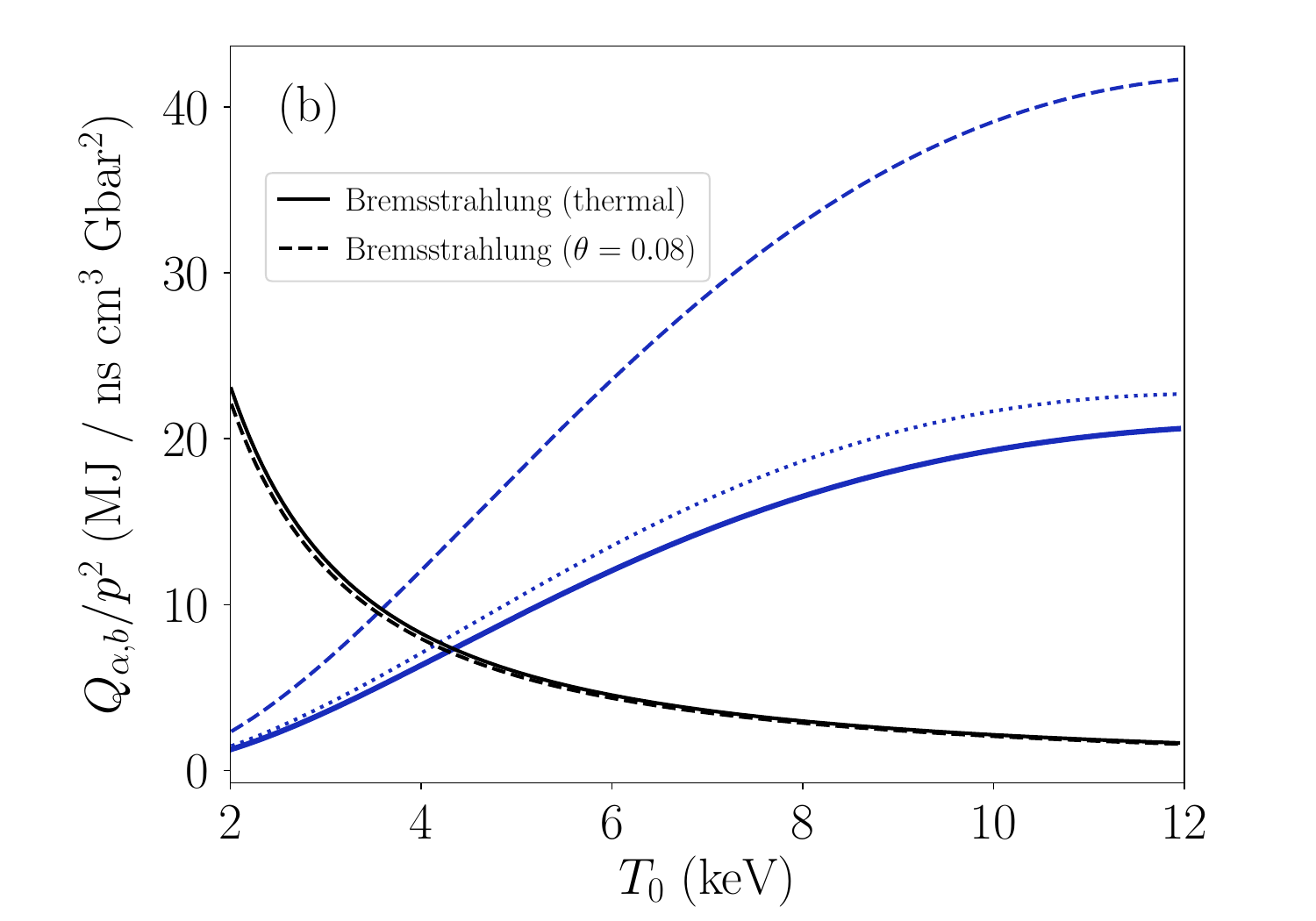}
    \label{fig_reactivity_T0_b}
  \end{subfigure}
  \caption{\justifying (a) Thermal\cite{Bosch_Hale_1992} and turbulence-enhanced reactivities\cite{Fetsch_Fisch_formula} for the DT [D(t,n)\textsuperscript{4}He], DD [D(d,p)T and D(d,n)\textsuperscript{3}He], and D\textsuperscript{3}He [\textsuperscript{3}He(d,p)T] reactions. The turbulent energy is $\int_0^\infty dk E(k) = T/4$, meaning $\theta = 1/12$ for DD and DT and $\theta = 1/15$ for D\textsuperscript{3}He. (b) DT alpha heating compared to bremsstrahlung at temperatures relevant to ICF (if all alpha particles are captured and all radiation escapes).}
  \label{fig_reactivity_T0}
\end{figure}

Suppose that some of the thermal energy of a fusion plasma is replaced with TKE. Holding the energy density and number density constant, the temperature must decrease, as shown schematically in Fig.~\ref{fig_TKE_sketch}. The flow is assumed to be isotropic and present only on scales smaller than the hot-spot radius. For simplicity, the flow is treated as divergence-free, although the density fluctuations associated with compressible turbulence would produce a small additional enhancement to reactivity.
The thermal pressure is replaced by an effective pressure, which includes both thermal and turbulent contributions \cite{Davidovits_Fisch_2019}. Let $p$ and $n$ take the same values as before, with $p$ now denoting the effective pressure, and let $\theta$ be the ratio of TKE to thermal energy so that ${p = 2nT(1+\theta)}$.
The effective temperature is $T_0=p/2n$ (in the absence of turbulence, $T_0 = T$). 
If the turbulent flow field is decomposed into solenoidal modes on various length scales, the contribution to the reactivity from each mode increases with its wavenumber $k$. 
According to Ref.~\citenum{Fetsch_Fisch_formula}, the reactivity is multiplied by the enhancement factor ${\Phi = 1 + 2\int dk E(k)G(k)}$, where $E(k)$ is the turbulent energy spectrum and $G(k)$ is a function derived in Ref.~\citenum{Fetsch_Fisch_formula}. 
Fig.~\ref{fig_reactivity_T0}a shows the DT, DD, and D\textsuperscript{3}He reactivities in thermal plasma and in ``turbulent'' plasma modeled by a shear single mode with kinetic energy per particle equal to $T$. The horizontal axis shows $T_0$; all points at the same $T_0$ have equal internal energy. Two cases are shown: one with a turbulent Knudsen number of $0.1$ and one with asymptotically large $k$; while not physically realizable, this latter case represents an upper bound on the reactivity enhancement for the TKE value shown. 
In Fig.~\ref{fig_reactivity_T0}b, the power due to alpha heating in equimolar DT plasma is compared to power radiated by bremsstrahlung in thermal $(\theta = 0)$ and turbulent $(\theta = 1/12)$ plasma, assuming ${f_\alpha = f_b = 1}$ for simplicity. The point where alpha heating exceeds radiative losses shifts to lower temperature due both to the increase in reactivity and to the decrease in radiation. Per Fig.~\ref{fig_reactivity_T0}a, a similar shift occurs for other reactions, which may enable ignition of D\textsuperscript{3}He and DD plasmas at temperatures more readily achievable in the laboratory. However, the remainder of this work considers only DT ignition.

\begin{figure}
\centering
\includegraphics[width=\columnwidth]{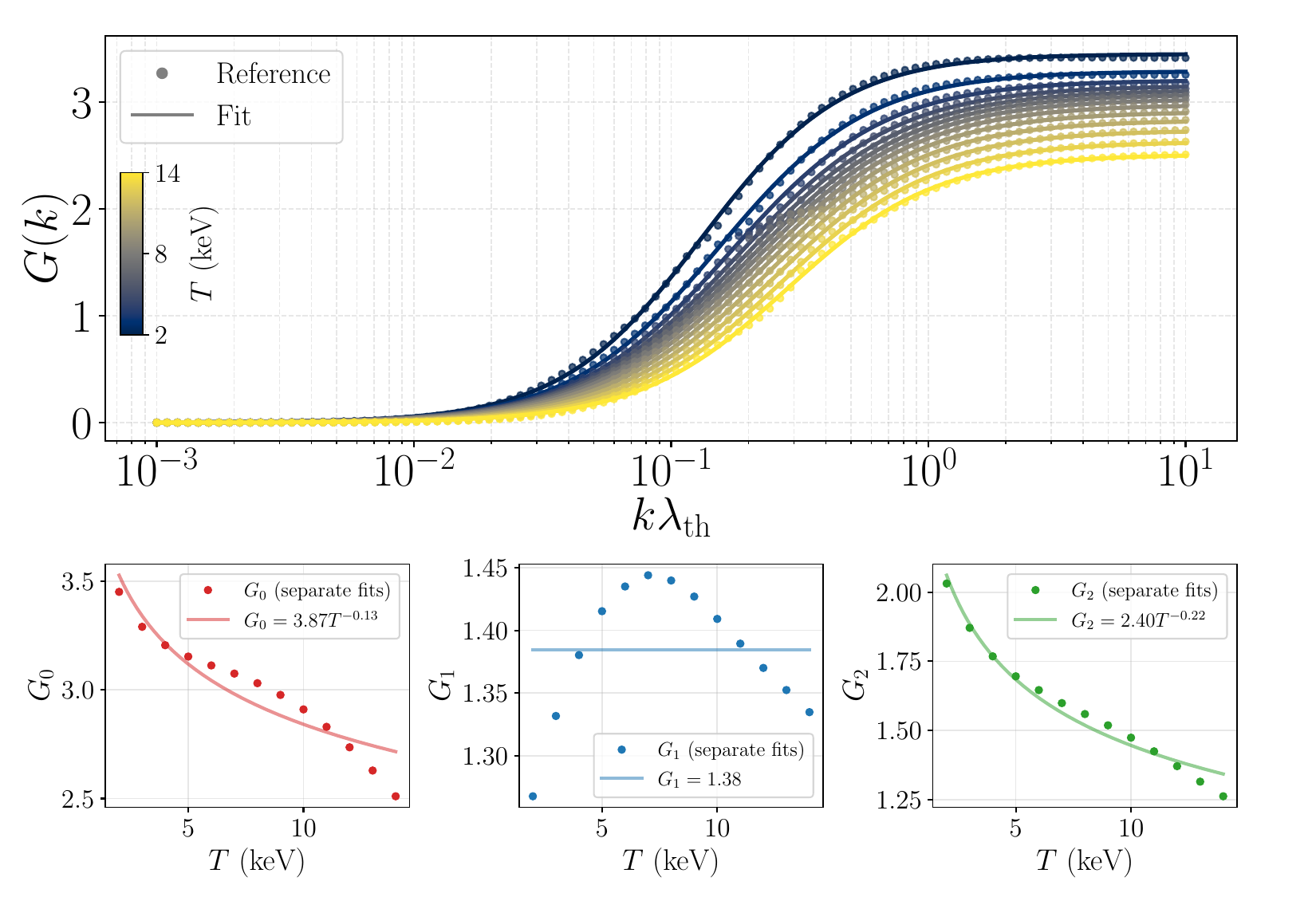}
\caption{\justifying Fit of $G(k)$ with parameters described in \eqref{eq_G_fit}. Reference data generated from the ``corrected utility function'' of Ref.~\citenum{Fetsch_Fisch_formula}.}
\label{fig_G_fit}
\end{figure} 

Accounting for the reactivity enhancement, the alpha-heating power can be written as ${Q_\alpha = f_\alpha p^2 \epsilon_\alpha S(T) (1 + H)}$. By the results of Ref.~\citenum{Fetsch_Fisch_formula}, $H$ takes the form
\begin{comment}
\begin{equation}
  \label{eq_H_def}
  H = \frac{6p^2}{\theta S(T)}\int_V d^3r  \frac{\avg{\sigma v}}{\wt T^3}\int_0^\infty dk \wt E(k) G(k),
\end{equation}
\end{comment}
\begin{equation}
  \label{eq_H_def}
  H = \frac{T}{\int_0^\infty dk E(k)}\left\langle 6 \int_0^\infty dk \frac{\wt E(k) G(k)}{\wt T} \right\rangle_\mr{react.}
\end{equation}
where the average is over the hotspot volume weighted by the local reaction rate (effectively neutron weighting). Note that ${\theta = \int_0^\infty dk E(k)/3T}$ for DT plasma. %(the factor of $\theta$ is pulled out of \eqref{eq_H_def} for convenience).
Here $\wt E(k, \bs r)$ is the local spectrum such that ${E(k) = \int_V d^3r \wt E(k, \bs r)/V}$. %Note that $G$ is also a function of the local density and temperature. 
Evaluating \eqref{eq_H_def} requires knowledge of the distribution of TKE throughout the hot spot, which is not specified in the model. However, because the reaction rate is peaked at the center of the hot spot, we approximate $H$ using the central parameters and assume that $E(k)$ is a good description of the energy spectrum at the center. Then \eqref{eq_H_def} simplifies to ${H = (6/\theta T)\int_0^\infty dk E(k) G(k)}$. 
To evaluate $G(k)$, the formula from Ref.~\citenum{Fetsch_Fisch_formula} is fit to a hyperbolic tangent with temperature-dependent parameters, \textit{viz.}
\begin{equation}
  \label{eq_G_fit}
  G(k) = \frac{G_0(T)}{2}\paren{1 + \tanh\paren{\frac{\ln(k\lambda_\mr{th}) - G_2(T)}{G_1}}}
\end{equation}
where $\lambda_\mr{th}$ is the mean free path of thermal ions (treated as one species whose mass is the average of the deuterium and tritium masses). The fit parameters are ${G_0 \approx 3.87 T^{-0.13}}$, ${G_1 \approx 1.38}$, and ${G_2 \approx 2.40T^{-0.22}}$, with $T$ in keV. 
The fit is shown in Fig.~\ref{fig_G_fit}; evidently, the shape of $G(k)$ is well captured by \eqref{eq_G_fit}. Note that $G$ depends on temperature and, through $\lambda_\mr{th}$, on density.

\begin{figure}[t]
\centering
\includegraphics[width=\columnwidth]{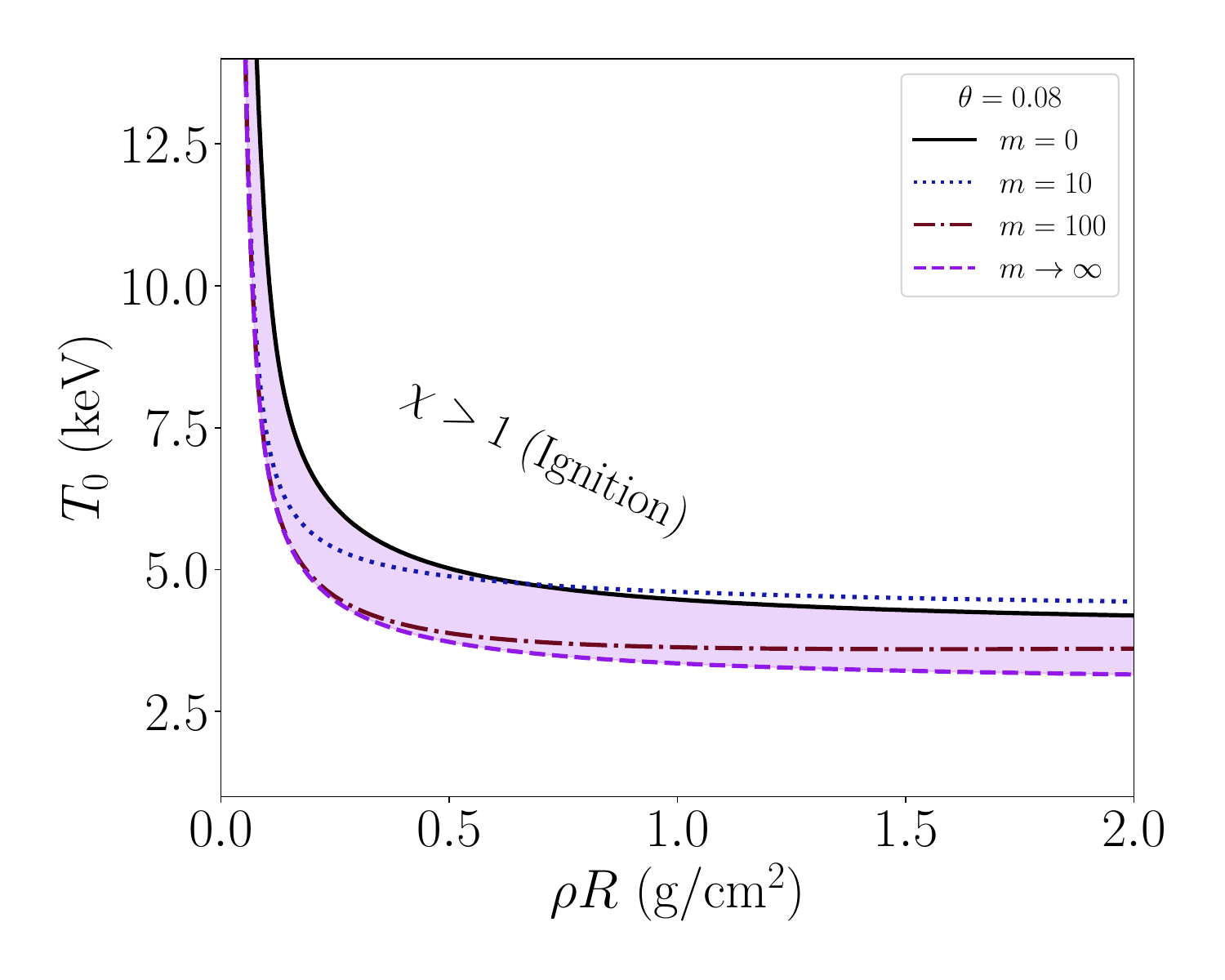}
\caption{\justifying Ignition contours $\chiTKE = 1$ for implosions with ${v_\mr{imp} = 400\text{km/s}}$ and perturbations of $\theta = 1/12$ and ``mode number'' $m$ such that ${k = 2\pi m/R}$. The $m=0$ case recovers the non-turbulent ignition parameter. The $m \to \infty$ case shows the maximum enhancement for this value of $\theta$. In the shaded region, where ${\chiTKE > 1 > \chi}$, ignition requires the enhancement of reactivity by turbulence.}
\label{fig_chi_TKE}
\end{figure}

Now, suppose that we introduce some turbulence into an ICF hot spot.  
The ignition condition becomes $\chiTKE > 1$, where the modified ignition parameter is
\begin{equation}
  \label{eq_chi_tke_def}
  \chi_\mr{turb} = \paren{\frac{f_\alpha \epsilon_\alpha S_0 T_0^{\omega-2}}{(1 + \theta)^{\omega}} (1+\theta H) - \frac{f_b B_0 T_0^{\beta-2}}{(1 + \theta)^{\beta}}(1 + \theta M)} \tauE p ,
\end{equation}
and $M$ estimates the increase in emission due to turbulent mixing of high-$Z$ material into the hot spot \cite{Patel_et_2020}. The form of \eqref{eq_chi_tke_def} is chosen to highlight the dominant temperature dependence in $\chiTKE$, but $f_\alpha$, $f_b$, and $\tauE$ may also vary with temperature and should be evaluated at ${T = T_0/(1+\theta)}$. 
In the rest of this work, we will take $M = 0$ for simplicity. This is reasonable if turbulence is driven near the center of the hot spot with little mixing at the boundary. We take $f_b = 0.3$ as a constant. For $f_\alpha$, we follow Ref.~\citenum{Lindl_et_2018} to write
\begin{equation}
  \label{eq_f_a_formual}
  f_\alpha = 
  \begin{cases}
    \frac{3}{2}\ell - \frac{4}{5}\ell^2, &  \ell \leq 1/2 \\
    1 - \frac{1}{4\ell} + \frac{1}{160\ell^3},  & \ell > 1/2
  \end{cases}
\end{equation}
where $\ell = 66.7(T^{-1.25} + 0.0082 )\rho R $ for $\rho R$ in ${\mr{g/cm}^2}$.

%TODO: cite Pak+ 2020 PRL for impact of turbulent mixing

The modified ignition condition is shown in Fig.~\ref{fig_chi_TKE} as contours in the space of hot-spot areal density $\rho R$ and turbulence-free temperature $T_0$. 
Each curve corresponds to turbulence on a scale ${k = 2\pi m/R}$, where $m$ is a mode number. This choice of variables is convenient because the curves then depend on $\rho R$ rather than on $\rho$ or $R$ individually; it also has physical significance if turbulence is driven by high-mode perturbations. 
All cases shown correspond to $\theta = 1/12$.
The $m=0$ case (solid black curve) recovers the non-turbulent ignition parameter $\chi$. In the $m=10$ case, ignition becomes easier in some regimes (low $\rho R$, high $T$) and harder in others (high $\rho R$, low $T$). The $m \to \infty$ case (dashed purple curve) shows the maximum enhancement for $\theta = 1/12$, corresponding to the saturation of $G$ at high $k$. At points in the shaded region, where ${\chiTKE > 1 > \chi}$, ignition is possible with TKE of $\theta = 1/12$ but impossible without it. 
The shift in the ignition contour happens for two reasons: at high $\rho R$, the contour shifts downward if the reactivity contribution from high-$k$ turbulence exceeds the decrease in reactivity due to reducing $T$ (at high $T$, the change in radiation plays a minor role). At low $\rho R$, reducing $T$ carries the additional benefit of reducing the alpha-particle mean free path, permitting ignition of a smaller hot spot.

\begin{figure}
  \centering
  \includegraphics[width=\columnwidth]{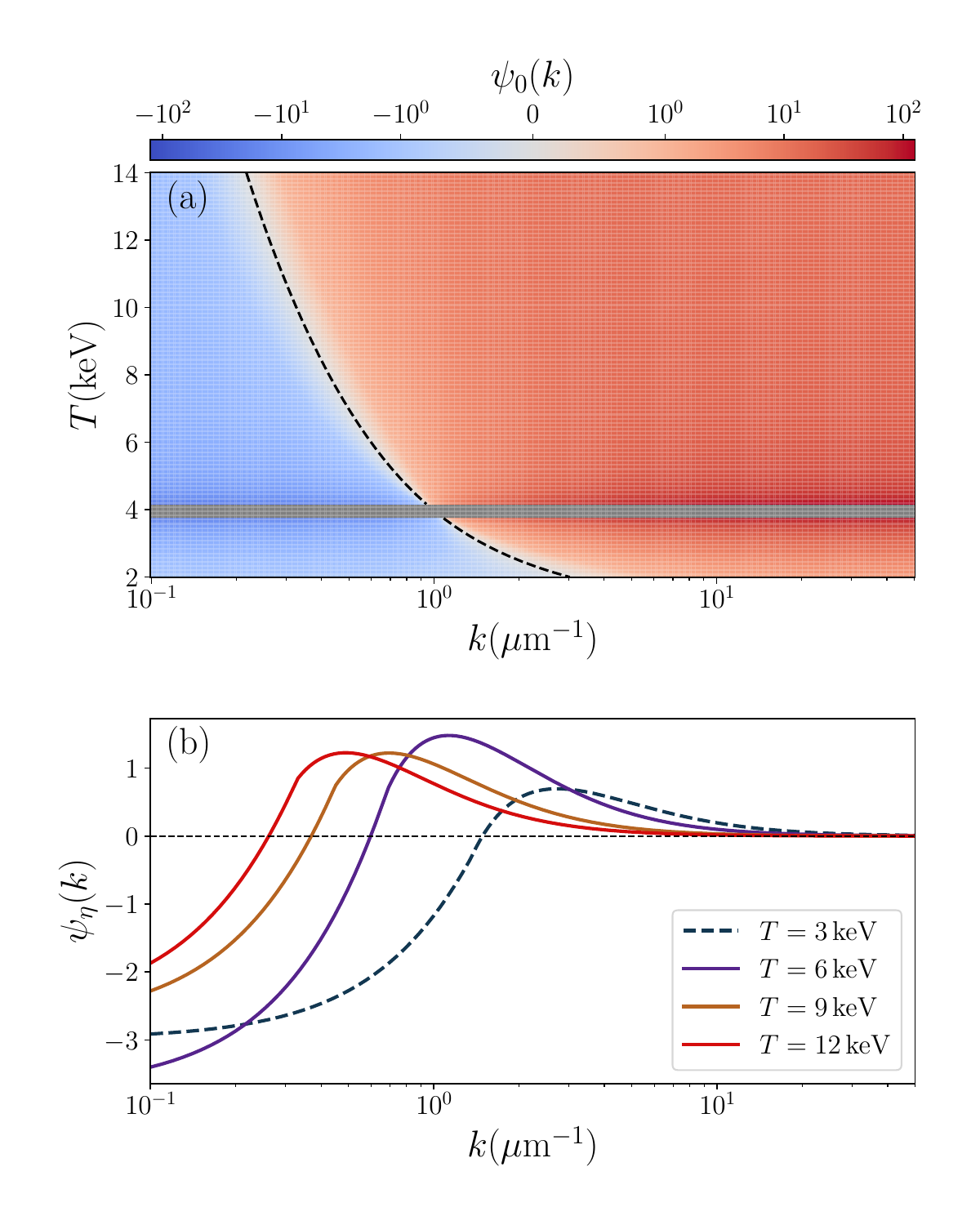}
  \caption{\justifying (a) Per-mode efficiency functions for a range of temperatures and wavenumbers at $\rho = 100~\mr{g/cm}^3$, $R = 50~\mu\text{m}$, and $\tauE = 50~\text{ps}$. (a) $\psi_0$ (without viscosity). The region near $\chi = 0$ where \eqref{eq_psi_k_def} breaks down is shaded in grey. (b) $\psi_\eta$ (with viscosity) for four values of $T$; the $3~\mr{keV}$ case is marked with a dashed line to highlight that $\chi < 0$.}
  \label{fig_psi_k}
\end{figure}

It is clear from Fig.~\ref{fig_chi_TKE} that turbulence is beneficial on some scales and detrimental on others. 
A key question for the design of ICF experiments is then: \textit{if hot-spot thermal energy is exchanged for TKE with a spectrum $E(k)$, does the resulting system ignite more easily?} 
To aid in answering this question, we quantify the improvement in $\chiTKE$ associated with an initial turbulent energy spectrum $E_0(k)$ at the start of the burn ($t = t_0$) by defining an efficiency function $\Psi[E_0]$ as
\begin{equation}
  \label{eq_Psi_def}
  \Psi[E_0] = \frac{\int_{t_0}^{t_f} dt \chiTKE}{\int_{t_0}^{t_f} dt \chi} - 1 ,
\end{equation}
where $t_f$ represents the end of the burn phase, and $\chiTKE$ depends on the energy spectrum, which evolves as
\begin{equation}
  \label{eq_E_evol_Gamma}
  E(k, t) = e^{\int_{t_0}^t dt^{\prime} \paren{\hat K - k^2 \eta}} E_0(k) ,
\end{equation}
where $\eta$ is the kinematic viscosity and $\hat K$ is a complicated operator accounting for the exchange of TKE between scales as well as amplification of TKE by compression\cite{Davidovits_Fisch_2017,Davidovits_Fisch_2019}. 
In general, \eqref{eq_E_evol_Gamma} requires simulations. However, one can seek a useful estimate by asking how $\chiTKE$ responds to a small amount of energy driven in a single mode $k$. % in the time before the energy of that mode is dissipated. 
For simplicity, we consider perturbations about a turbulence-free state ${(\theta = 0)}$. This per-mode efficiency function $\psi_0(k)$ is defined as
\begin{equation}
  \label{eq_psi_k_def}
  \psi_0(k) = \frac{1}{|\chi|} \paren{\frac{\partial \chiTKE}{\partial \theta}}_{p, \rho} \Bigg |_{\hat E(k') = \delta(k-k'), \theta = 0} ,
\end{equation}
which, using \eqref{eq_chi_tke_def}, takes the following simple form:
\begin{equation}
  \label{eq_psi_k_soln}
  \psi_0(k) =\mr{sgn}(\chi)\paren{\beta + \phi_b + \frac{6G(k) + \beta + \phi_b - \omega - \phi_\alpha - M}{\paren{1 - \paren{\frac{T}{\Tign}}^{\beta - \omega}}} } ,
\end{equation}
where $\mathrm{sgn}$ takes the sign ($-1$ or $1$) of its argument and ${\phi_{\alpha/b} = -Td(\ln f_{\alpha/b})/dT}$. % record the temperature dependence of the capture and loss fractions for alpha particles and radiation. 
The behavior of $\psi_0(k,T)$ is shown in Fig.~\ref{fig_psi_k}a in a hot spot with ${\rho = 100~\mr{g/cm}^3}$, ${R = 50~\mu\text{m}}$, and ${\tauE = 50~\text{ps}}$. To the right of the $\psi_0 = 0$ contour (black dashed line), exchanging some thermal energy for TKE is advantageous. This advantage increases dramatically with $k$ on sub-micron scales, reaching values around $\psi_0 \sim 100$.
%At $6~\mr{keV}$, $\psi_0 > 0$ for all $k$ shown. For $9~\mr{keV}$ and $12~\mr{keV}$, $\psi_0$ becomes positive at TKE length scales slightly shorter than $1~\mu\text{m}$. In all cases, $\psi_0$ increases rapdily with $k$ on sub-micron scales.

On these short length scales, viscosity plays an important role, rapidly dissipating TKE. Thus, the time-integrated impact of TKE in high-$k$ modes is dampened because energy resides for a shorter time in these modes before being converted to thermal energy. 
To account for this tradeoff, we define a \textit{viscous} single-mode efficiency function ${\psi_\eta(k) = (\tau_\eta/\tauE)\psi_0(k)}$, where the viscous dissipation time is ${\tau_\eta = \min\paren{1/k^2\eta, \tauE}}$. 
While this definition elides several important effects, for example the nonlinear feedback by thermonuclear instability, $\psi_\eta$ nevertheless provides a useful estimate. 
Fig.~\ref{fig_psi_k}b shows $\psi_\eta(k)$ at several temperatures for the same parameters as in Fig.~\ref{fig_psi_k}a. Due to the $k^{-2}$ dependence of $\tau_\eta$, $\psi_\eta(k)$ has a peak at some $k_\mr{opt}$, which depends on temperature. Within the model of viscous dissipation adopted here, $k_\mr{opt}$ is the wavenumber at which TKE produces the largest time-integrated increase in $\chiTKE$. 
In optimizing ICF implosions, it is likely desirable to drive flows on length scales around $1/k_\mr{opt}$. Such driving could be accomplished by embedding perturbations designed to become Rayleigh-Taylor or Richtmyer-Meshkov unstable at stagnation. 
In structured materials, such as wetted foams, the collapse of voids during compression generates flows and ion-kinetic effects\cite{Yin_Albright_Vold_Nystrom_Bird_Bowers_2019}, suggesting that engineered foams could be used to drive turbulence on advantageous scales.

%For clarity, we offer a simple numerical formula for $\psi_0$ using formulas derived above. For a plasma above $T_\mr{ign}$, assuming that $X \approx 0$, $\ell \gg 1$ such that $f_\alpha \approx 1$, and $f_b$ is nearly constant, \eqref{eq_psi_k_soln} simplifies to
%\begin{equation}
%  \label{eq_psi_k_numerical}
 % \psi_0(k) \approx \half + \frac{1}{1 - (T/4.3f_b)^{-2.76}} \paren{-2.76 + 23.22T^{-0.13} \frac{}{}}
%\end{equation}

In this work, we derived a Lawson-like ignition criterion, ${\chiTKE > 1}$, for implosions generalized to the case of compressing turbulent plasma \eqref{eq_chi_tke_def}. Comparison to an ignition parameter based on the GLC of Betti \textit{et al.}\cite{Betti_et_2010} demonstrates a new regime (Fig.~\ref{fig_chi_TKE}) in which ignition is possible only by driving small-scale turbulence. An approximate efficiency function $\psi_\eta$ for driving turbulence demonstrates (Fig.~\ref{fig_psi_k}) that ICF experiments would benefit from micron-scale turbulence deep inside the hot spot. Turbulence on long length scales is, of course, still deleterious. 
This work used a 1D hot-spot model to focus on the new physics of the \sfe, but the conclusions are readily generalized to more realistic asymmetric hot spots. In fact, $\psi_0$ and $\psi_\eta$ are independent of hot-spot geometry, meaning that the conclusions about optimal turbulence length scales are quite general.

\acknowledgments
This work was supported by the Center for Magnetic Acceleration, Compression, and Heating (MACH), part of the U.S. DOE-NNSA Stewardship Science Academic Alliances Program under Cooperative Agreement DE-NA0004148.

\bibliography{turb_lawson_crit}

\providecommand{\noopsort}[1]{}\providecommand{\singleletter}[1]{#1}%
\begin{thebibliography}{19}%
\makeatletter
\providecommand \@ifxundefined [1]{%
 \@ifx{#1\undefined}
}%
\providecommand \@ifnum [1]{%
 \ifnum #1\expandafter \@firstoftwo
 \else \expandafter \@secondoftwo
 \fi
}%
\providecommand \@ifx [1]{%
 \ifx #1\expandafter \@firstoftwo
 \else \expandafter \@secondoftwo
 \fi
}%
\providecommand \natexlab [1]{#1}%
\providecommand \enquote  [1]{``#1''}%
\providecommand \bibnamefont  [1]{#1}%
\providecommand \bibfnamefont [1]{#1}%
\providecommand \citenamefont [1]{#1}%
\providecommand \href@noop [0]{\@secondoftwo}%
\providecommand \href [0]{\begingroup \@sanitize@url \@href}%
\providecommand \@href[1]{\@@startlink{#1}\@@href}%
\providecommand \@@href[1]{\endgroup#1\@@endlink}%
\providecommand \@sanitize@url [0]{\catcode `\\12\catcode `\$12\catcode `\&12\catcode `\#12\catcode `\^12\catcode `\_12\catcode `\%12\relax}%
\providecommand \@@startlink[1]{}%
\providecommand \@@endlink[0]{}%
\providecommand \url  [0]{\begingroup\@sanitize@url \@url }%
\providecommand \@url [1]{\endgroup\@href {#1}{\urlprefix }}%
\providecommand \urlprefix  [0]{URL }%
\providecommand \Eprint [0]{\href }%
\providecommand \doibase [0]{https://doi.org/}%
\providecommand \selectlanguage [0]{\@gobble}%
\providecommand \bibinfo  [0]{\@secondoftwo}%
\providecommand \bibfield  [0]{\@secondoftwo}%
\providecommand \translation [1]{[#1]}%
\providecommand \BibitemOpen [0]{}%
\providecommand \bibitemStop [0]{}%
\providecommand \bibitemNoStop [0]{.\EOS\space}%
\providecommand \EOS [0]{\spacefactor3000\relax}%
\providecommand \BibitemShut  [1]{\csname bibitem#1\endcsname}%
\let\auto@bib@innerbib\@empty
%</preamble>
\bibitem [{\citenamefont {Lawson}(1957)}]{Lawson_1957}%
  \BibitemOpen
  \bibfield  {author} {\bibinfo {author} {\bibfnamefont {J.~D.}\ \bibnamefont {Lawson}},\ }\href {https://doi.org/10.1088/0370-1301/70/1/303} {\bibfield  {journal} {\bibinfo  {journal} {Proceedings of the Physical Society. Section B}\ }\textbf {\bibinfo {volume} {70}},\ \bibinfo {pages} {6} (\bibinfo {year} {1957})}\BibitemShut {NoStop}%
\bibitem [{\citenamefont {Lindl}\ \emph {et~al.}(2018)\citenamefont {Lindl}, \citenamefont {Haan}, \citenamefont {Landen}, \citenamefont {Christopherson},\ and\ \citenamefont {Betti}}]{Lindl_et_2018}%
  \BibitemOpen
  \bibfield  {author} {\bibinfo {author} {\bibfnamefont {J.~D.}\ \bibnamefont {Lindl}}, \bibinfo {author} {\bibfnamefont {S.~W.}\ \bibnamefont {Haan}}, \bibinfo {author} {\bibfnamefont {O.~L.}\ \bibnamefont {Landen}}, \bibinfo {author} {\bibfnamefont {A.~R.}\ \bibnamefont {Christopherson}},\ and\ \bibinfo {author} {\bibfnamefont {R.}~\bibnamefont {Betti}},\ }\href {https://doi.org/10.1063/1.5049595} {\bibfield  {journal} {\bibinfo  {journal} {Physics of Plasmas}\ }\textbf {\bibinfo {volume} {25}},\ \bibinfo {pages} {122704} (\bibinfo {year} {2018})}\BibitemShut {NoStop}%
\bibitem [{\citenamefont {Zhou}\ and\ \citenamefont {Betti}(2008)}]{Zhou_Betti_2008}%
  \BibitemOpen
  \bibfield  {author} {\bibinfo {author} {\bibfnamefont {C.~D.}\ \bibnamefont {Zhou}}\ and\ \bibinfo {author} {\bibfnamefont {R.}~\bibnamefont {Betti}},\ }\href {https://doi.org/10.1063/1.2998604} {\bibfield  {journal} {\bibinfo  {journal} {Physics of Plasmas}\ }\textbf {\bibinfo {volume} {15}},\ \bibinfo {pages} {102707} (\bibinfo {year} {2008})}\BibitemShut {NoStop}%
\bibitem [{\citenamefont {Betti}\ \emph {et~al.}(2010)\citenamefont {Betti}, \citenamefont {Chang}, \citenamefont {Spears}, \citenamefont {Anderson}, \citenamefont {Edwards}, \citenamefont {Fatenejad}, \citenamefont {Lindl}, \citenamefont {McCrory}, \citenamefont {Nora},\ and\ \citenamefont {Shvarts}}]{Betti_et_2010}%
  \BibitemOpen
  \bibfield  {author} {\bibinfo {author} {\bibfnamefont {R.}~\bibnamefont {Betti}}, \bibinfo {author} {\bibfnamefont {P.~Y.}\ \bibnamefont {Chang}}, \bibinfo {author} {\bibfnamefont {B.~K.}\ \bibnamefont {Spears}}, \bibinfo {author} {\bibfnamefont {K.~S.}\ \bibnamefont {Anderson}}, \bibinfo {author} {\bibfnamefont {J.}~\bibnamefont {Edwards}}, \bibinfo {author} {\bibfnamefont {M.}~\bibnamefont {Fatenejad}}, \bibinfo {author} {\bibfnamefont {J.~D.}\ \bibnamefont {Lindl}}, \bibinfo {author} {\bibfnamefont {R.~L.}\ \bibnamefont {McCrory}}, \bibinfo {author} {\bibfnamefont {R.}~\bibnamefont {Nora}},\ and\ \bibinfo {author} {\bibfnamefont {D.}~\bibnamefont {Shvarts}},\ }\bibfield  {journal} {\bibinfo  {journal} {Physics of Plasmas}\ }\textbf {\bibinfo {volume} {17}},\ \href {https://doi.org/10.1063/1.3380857} {10.1063/1.3380857} (\bibinfo {year} {2010})\BibitemShut {NoStop}%
\bibitem [{\citenamefont {Christopherson}\ \emph {et~al.}(2018)\citenamefont {Christopherson}, \citenamefont {Betti}, \citenamefont {Bose}, \citenamefont {Howard}, \citenamefont {Woo}, \citenamefont {Campbell}, \citenamefont {Sanz},\ and\ \citenamefont {Spears}}]{Christopherson_et_2018}%
  \BibitemOpen
  \bibfield  {author} {\bibinfo {author} {\bibfnamefont {A.~R.}\ \bibnamefont {Christopherson}}, \bibinfo {author} {\bibfnamefont {R.}~\bibnamefont {Betti}}, \bibinfo {author} {\bibfnamefont {A.}~\bibnamefont {Bose}}, \bibinfo {author} {\bibfnamefont {J.}~\bibnamefont {Howard}}, \bibinfo {author} {\bibfnamefont {K.~M.}\ \bibnamefont {Woo}}, \bibinfo {author} {\bibfnamefont {E.~M.}\ \bibnamefont {Campbell}}, \bibinfo {author} {\bibfnamefont {J.}~\bibnamefont {Sanz}},\ and\ \bibinfo {author} {\bibfnamefont {B.~K.}\ \bibnamefont {Spears}},\ }\href {https://doi.org/10.1063/1.4991405} {\bibfield  {journal} {\bibinfo  {journal} {Physics of Plasmas}\ }\textbf {\bibinfo {volume} {25}},\ \bibinfo {pages} {012703} (\bibinfo {year} {2018})}\BibitemShut {NoStop}%
\bibitem [{\citenamefont {Zylstra}\ \emph {et~al.}(2022)\citenamefont {Zylstra}, \citenamefont {Kritcher}, \citenamefont {Hurricane}, \citenamefont {Callahan}, \citenamefont {Ralph}, \citenamefont {Casey}, \citenamefont {Pak}, \citenamefont {Landen}, \citenamefont {Bachmann}, \citenamefont {Baker} \emph {et~al.}}]{Zylstra_et_2022}%
  \BibitemOpen
  \bibfield  {author} {\bibinfo {author} {\bibfnamefont {A.~B.}\ \bibnamefont {Zylstra}}, \bibinfo {author} {\bibfnamefont {A.~L.}\ \bibnamefont {Kritcher}}, \bibinfo {author} {\bibfnamefont {O.~A.}\ \bibnamefont {Hurricane}}, \bibinfo {author} {\bibfnamefont {D.~A.}\ \bibnamefont {Callahan}}, \bibinfo {author} {\bibfnamefont {J.~E.}\ \bibnamefont {Ralph}}, \bibinfo {author} {\bibfnamefont {D.~T.}\ \bibnamefont {Casey}}, \bibinfo {author} {\bibfnamefont {A.}~\bibnamefont {Pak}}, \bibinfo {author} {\bibfnamefont {O.~L.}\ \bibnamefont {Landen}}, \bibinfo {author} {\bibfnamefont {B.}~\bibnamefont {Bachmann}}, \bibinfo {author} {\bibfnamefont {K.~L.}\ \bibnamefont {Baker}}, \emph {et~al.},\ }\href {https://doi.org/10.1103/PhysRevE.106.025202} {\bibfield  {journal} {\bibinfo  {journal} {Physical Review E}\ }\textbf {\bibinfo {volume} {106}},\ \bibinfo {pages} {025202} (\bibinfo {year} {2022})}\BibitemShut {NoStop}%
\bibitem [{\citenamefont {Hurricane}\ \emph {et~al.}(2021)\citenamefont {Hurricane}, \citenamefont {Maclaren}, \citenamefont {Rosen}, \citenamefont {Hammer}, \citenamefont {Springer},\ and\ \citenamefont {Betti}}]{Hurricane_et_2021}%
  \BibitemOpen
  \bibfield  {author} {\bibinfo {author} {\bibfnamefont {O.~A.}\ \bibnamefont {Hurricane}}, \bibinfo {author} {\bibfnamefont {S.~A.}\ \bibnamefont {Maclaren}}, \bibinfo {author} {\bibfnamefont {M.~D.}\ \bibnamefont {Rosen}}, \bibinfo {author} {\bibfnamefont {J.~H.}\ \bibnamefont {Hammer}}, \bibinfo {author} {\bibfnamefont {P.~T.}\ \bibnamefont {Springer}},\ and\ \bibinfo {author} {\bibfnamefont {R.}~\bibnamefont {Betti}},\ }\href {https://doi.org/10.1063/5.0035583} {\bibfield  {journal} {\bibinfo  {journal} {Physics of Plasmas}\ }\textbf {\bibinfo {volume} {28}},\ \bibinfo {pages} {022704} (\bibinfo {year} {2021})}\BibitemShut {NoStop}%
\bibitem [{\citenamefont {Hurricane}\ \emph {et~al.}(2024)\citenamefont {Hurricane}, \citenamefont {Allen}, \citenamefont {Bachmann}, \citenamefont {Baker}, \citenamefont {Baxamusa}, \citenamefont {Bhandarkar}, \citenamefont {Biener}, \citenamefont {Bionta}, \citenamefont {Braun}, \citenamefont {Briggs} \emph {et~al.}}]{Hurricane_et_2024}%
  \BibitemOpen
  \bibfield  {author} {\bibinfo {author} {\bibfnamefont {O.~A.}\ \bibnamefont {Hurricane}}, \bibinfo {author} {\bibfnamefont {A.}~\bibnamefont {Allen}}, \bibinfo {author} {\bibfnamefont {B.~L.}\ \bibnamefont {Bachmann}}, \bibinfo {author} {\bibfnamefont {K.~L.}\ \bibnamefont {Baker}}, \bibinfo {author} {\bibfnamefont {S.}~\bibnamefont {Baxamusa}}, \bibinfo {author} {\bibfnamefont {S.~D.}\ \bibnamefont {Bhandarkar}}, \bibinfo {author} {\bibfnamefont {J.}~\bibnamefont {Biener}}, \bibinfo {author} {\bibfnamefont {S.~R.~M.}\ \bibnamefont {Bionta}}, \bibinfo {author} {\bibfnamefont {T.}~\bibnamefont {Braun}}, \bibinfo {author} {\bibfnamefont {T.}~\bibnamefont {Briggs}}, \emph {et~al.},\ }\href {https://doi.org/10.1088/1361-6587/ad994f} {\bibfield  {journal} {\bibinfo  {journal} {Plasma Physics and Controlled Fusion}\ }\textbf {\bibinfo {volume} {67}},\ \bibinfo {pages} {015019} (\bibinfo {year} {2024})}\BibitemShut {NoStop}%
\bibitem [{\citenamefont {Christopherson}\ \emph {et~al.}(2020)\citenamefont {Christopherson}, \citenamefont {Betti}, \citenamefont {Miller}, \citenamefont {Gopalaswamy}, \citenamefont {Mannion},\ and\ \citenamefont {Cao}}]{Christopherson_et_2020}%
  \BibitemOpen
  \bibfield  {author} {\bibinfo {author} {\bibfnamefont {A.~R.}\ \bibnamefont {Christopherson}}, \bibinfo {author} {\bibfnamefont {R.}~\bibnamefont {Betti}}, \bibinfo {author} {\bibfnamefont {S.}~\bibnamefont {Miller}}, \bibinfo {author} {\bibfnamefont {V.}~\bibnamefont {Gopalaswamy}}, \bibinfo {author} {\bibfnamefont {O.~M.}\ \bibnamefont {Mannion}},\ and\ \bibinfo {author} {\bibfnamefont {D.}~\bibnamefont {Cao}},\ }\href {https://doi.org/10.1063/1.5143889} {\bibfield  {journal} {\bibinfo  {journal} {Physics of Plasmas}\ }\textbf {\bibinfo {volume} {27}},\ \bibinfo {pages} {052708} (\bibinfo {year} {2020})}\BibitemShut {NoStop}%
\bibitem [{\citenamefont {{The Indirect Drive ICF Collaboration}}\ \emph {et~al.}(2008)\citenamefont {{The Indirect Drive ICF Collaboration}}, \citenamefont {Abu-Shawareb} \emph {et~al.}}]{ICF_collab_2024}%
  \BibitemOpen
  \bibfield  {author} {\bibinfo {author} {\bibnamefont {{The Indirect Drive ICF Collaboration}}}, \bibinfo {author} {\bibfnamefont {H.}~\bibnamefont {Abu-Shawareb}}, \emph {et~al.},\ }\href {https://doi.org/10.1103/PhysRevLett.132.065102} {\bibfield  {journal} {\bibinfo  {journal} {Physical Review Letters}\ }\textbf {\bibinfo {volume} {132}},\ \bibinfo {pages} {102707} (\bibinfo {year} {2008})}\BibitemShut {NoStop}%
\bibitem [{\citenamefont {Kritcher}\ \emph {et~al.}(2024)\citenamefont {Kritcher}, \citenamefont {Zylstra}, \citenamefont {Weber}, \citenamefont {Hurricane}, \citenamefont {Callahan}, \citenamefont {Clark}, \citenamefont {Divol}, \citenamefont {Hinkel}, \citenamefont {Humbird}, \citenamefont {Jones} \emph {et~al.}}]{Kritcher_et_2024}%
  \BibitemOpen
  \bibfield  {author} {\bibinfo {author} {\bibfnamefont {A.~L.}\ \bibnamefont {Kritcher}}, \bibinfo {author} {\bibfnamefont {A.~B.}\ \bibnamefont {Zylstra}}, \bibinfo {author} {\bibfnamefont {C.~R.}\ \bibnamefont {Weber}}, \bibinfo {author} {\bibfnamefont {O.~A.}\ \bibnamefont {Hurricane}}, \bibinfo {author} {\bibfnamefont {D.~A.}\ \bibnamefont {Callahan}}, \bibinfo {author} {\bibfnamefont {D.~S.}\ \bibnamefont {Clark}}, \bibinfo {author} {\bibfnamefont {L.}~\bibnamefont {Divol}}, \bibinfo {author} {\bibfnamefont {D.~E.}\ \bibnamefont {Hinkel}}, \bibinfo {author} {\bibfnamefont {K.}~\bibnamefont {Humbird}}, \bibinfo {author} {\bibfnamefont {O.}~\bibnamefont {Jones}}, \emph {et~al.},\ }\href {https://doi.org/10.1103/PhysRevE.109.025204} {\bibfield  {journal} {\bibinfo  {journal} {Physical Review E}\ }\textbf {\bibinfo {volume} {109}},\ \bibinfo {pages} {025204} (\bibinfo {year} {2024})}\BibitemShut {NoStop}%
\bibitem [{\citenamefont {Rosen}(2024)}]{Rosen_2024}%
  \BibitemOpen
  \bibfield  {author} {\bibinfo {author} {\bibfnamefont {M.~D.}\ \bibnamefont {Rosen}},\ }\href {https://doi.org/10.1063/5.0221005} {\bibfield  {journal} {\bibinfo  {journal} {Physics of Plasmas}\ }\textbf {\bibinfo {volume} {31}},\ \bibinfo {pages} {090501} (\bibinfo {year} {2024})}\BibitemShut {NoStop}%
\bibitem [{\citenamefont {Fetsch}\ and\ \citenamefont {Fisch}(2024)}]{Fetsch_Fisch_2024}%
  \BibitemOpen
  \bibfield  {author} {\bibinfo {author} {\bibfnamefont {H.}~\bibnamefont {Fetsch}}\ and\ \bibinfo {author} {\bibfnamefont {N.~J.}\ \bibnamefont {Fisch}},\ }\bibfield  {journal} {\bibinfo  {journal} {arXiv preprint}\ }\href {https://doi.org/10.48550/arXiv.2410.03590} {10.48550/arXiv.2410.03590} (\bibinfo {year} {2024}),\ \bibinfo {note} {arXiv:2410.03590 [physics]}\BibitemShut {NoStop}%
\bibitem [{\citenamefont {Fetsch}\ and\ \citenamefont {Fisch}(2025)}]{Fetsch_Fisch_formula}%
  \BibitemOpen
  \bibfield  {author} {\bibinfo {author} {\bibfnamefont {H.}~\bibnamefont {Fetsch}}\ and\ \bibinfo {author} {\bibfnamefont {N.~J.}\ \bibnamefont {Fisch}},\ }\bibfield  {journal} {\bibinfo  {journal} {arXiv preprint}\ }\href {https://doi.org/10.48550/arXiv.2506.13711} {10.48550/arXiv.2506.13711} (\bibinfo {year} {2025}),\ \bibinfo {note} {arXiv:2506.13711 [physics]}\BibitemShut {NoStop}%
\bibitem [{\citenamefont {Bosch}\ and\ \citenamefont {Hale}(1992)}]{Bosch_Hale_1992}%
  \BibitemOpen
  \bibfield  {author} {\bibinfo {author} {\bibfnamefont {H.-S.}\ \bibnamefont {Bosch}}\ and\ \bibinfo {author} {\bibfnamefont {G.~M.}\ \bibnamefont {Hale}},\ }\href {https://doi.org/10.1088/0029-5515/32/4/I07} {\bibfield  {journal} {\bibinfo  {journal} {Nuclear Fusion}\ }\textbf {\bibinfo {volume} {32}},\ \bibinfo {pages} {611} (\bibinfo {year} {1992})}\BibitemShut {NoStop}%
\bibitem [{\citenamefont {Davidovits}\ and\ \citenamefont {Fisch}(2019)}]{Davidovits_Fisch_2019}%
  \BibitemOpen
  \bibfield  {author} {\bibinfo {author} {\bibfnamefont {S.}~\bibnamefont {Davidovits}}\ and\ \bibinfo {author} {\bibfnamefont {N.~J.}\ \bibnamefont {Fisch}},\ }\href {https://doi.org/10.1063/1.5098790} {\bibfield  {journal} {\bibinfo  {journal} {Physics of Plasmas}\ }\textbf {\bibinfo {volume} {26}},\ \bibinfo {pages} {062709} (\bibinfo {year} {2019})}\BibitemShut {NoStop}%
\bibitem [{\citenamefont {Patel}\ \emph {et~al.}(2020)\citenamefont {Patel}, \citenamefont {Springer}, \citenamefont {Weber}, \citenamefont {Jarrott}, \citenamefont {Hurricane}, \citenamefont {Bachmann}, \citenamefont {Baker}, \citenamefont {Berzak~Hopkins}, \citenamefont {Callahan}, \citenamefont {Casey} \emph {et~al.}}]{Patel_et_2020}%
  \BibitemOpen
  \bibfield  {author} {\bibinfo {author} {\bibfnamefont {P.~K.}\ \bibnamefont {Patel}}, \bibinfo {author} {\bibfnamefont {P.~T.}\ \bibnamefont {Springer}}, \bibinfo {author} {\bibfnamefont {C.~R.}\ \bibnamefont {Weber}}, \bibinfo {author} {\bibfnamefont {L.~C.}\ \bibnamefont {Jarrott}}, \bibinfo {author} {\bibfnamefont {O.~A.}\ \bibnamefont {Hurricane}}, \bibinfo {author} {\bibfnamefont {B.}~\bibnamefont {Bachmann}}, \bibinfo {author} {\bibfnamefont {K.~L.}\ \bibnamefont {Baker}}, \bibinfo {author} {\bibfnamefont {L.~F.}\ \bibnamefont {Berzak~Hopkins}}, \bibinfo {author} {\bibfnamefont {D.~A.}\ \bibnamefont {Callahan}}, \bibinfo {author} {\bibfnamefont {D.~T.}\ \bibnamefont {Casey}}, \emph {et~al.},\ }\href {https://doi.org/10.1063/5.0003298} {\bibfield  {journal} {\bibinfo  {journal} {Physics of Plasmas}\ }\textbf {\bibinfo {volume} {27}},\ \bibinfo {pages} {050901} (\bibinfo {year} {2020})}\BibitemShut {NoStop}%
\bibitem [{\citenamefont {Davidovits}\ and\ \citenamefont {Fisch}(2017)}]{Davidovits_Fisch_2017}%
  \BibitemOpen
  \bibfield  {author} {\bibinfo {author} {\bibfnamefont {S.}~\bibnamefont {Davidovits}}\ and\ \bibinfo {author} {\bibfnamefont {N.~J.}\ \bibnamefont {Fisch}},\ }\href {https://doi.org/10.3847/1538-4357/aa619f} {\bibfield  {journal} {\bibinfo  {journal} {The Astrophysical Journal}\ }\textbf {\bibinfo {volume} {838}},\ \bibinfo {pages} {118} (\bibinfo {year} {2017})}\BibitemShut {NoStop}%
\bibitem [{\citenamefont {Yin}\ \emph {et~al.}(2019)\citenamefont {Yin}, \citenamefont {Albright}, \citenamefont {Vold}, \citenamefont {Nystrom}, \citenamefont {Bird},\ and\ \citenamefont {Bowers}}]{Yin_Albright_Vold_Nystrom_Bird_Bowers_2019}%
  \BibitemOpen
  \bibfield  {author} {\bibinfo {author} {\bibfnamefont {L.}~\bibnamefont {Yin}}, \bibinfo {author} {\bibfnamefont {B.~J.}\ \bibnamefont {Albright}}, \bibinfo {author} {\bibfnamefont {E.~L.}\ \bibnamefont {Vold}}, \bibinfo {author} {\bibfnamefont {W.~D.}\ \bibnamefont {Nystrom}}, \bibinfo {author} {\bibfnamefont {R.~F.}\ \bibnamefont {Bird}},\ and\ \bibinfo {author} {\bibfnamefont {K.~J.}\ \bibnamefont {Bowers}},\ }\href {https://doi.org/10.1063/1.5109257} {\bibfield  {journal} {\bibinfo  {journal} {Physics of Plasmas}\ }\textbf {\bibinfo {volume} {26}},\ \bibinfo {pages} {062302} (\bibinfo {year} {2019})}\BibitemShut {NoStop}%
\end{thebibliography}%

\end{document}